# Tell Machine Learning Potentials What They are Needed for: Simulation-Oriented Training Exemplified for Glycine

Fuchun Ge,[1][†] Ran Wang,[1][†] Chen Qu,[2] Peikun Zheng,[1,3] Apurba Nandi,[4,5]
Riccardo Conte,[6] Paul L. Houston,[7] Joel M. Bowman,[4][*] Pavlo O. Dral[1][*]

[1]*State Key Laboratory of Physical Chemistry of Solid Surfaces, College of Chemistry and Chemical Engineering, Fujian Provincial Key Laboratory of Theoretical and Computational Chemistry, and Innovation Laboratory for Sciences and Technologies of Energy Materials of Fujian Province (IKKEM), Xiamen University, Xiamen, Fujian 361005, China*

[2]*Independent Researcher, Toronto, Ontario M9B0E3, Canada*

[3]Present address: *Department of Chemistry, Carnegie Mellon University, Pittsburgh, Pennsylvania 15213, United States*

[4]*Department of Chemistry and Cherry L. Emerson Center for Scientific Computation, Emory University, Atlanta, Georgia, 30322, United States*

[5]*Department of Physics and Materials Science, University of Luxembourg, Luxembourg City L-1511, Luxembourg*

[6]*Dipartimento di Chimica, Università degli Studi di Milano, via Golgi 19, 20133 Milano, Italy*

[7]*Department of Chemistry and Chemical Biology, Cornell University, Ithaca, New York 14853, United States; Department of Chemistry and Biochemistry, Georgia Institute of Technology, Atlanta, Georgia 30332, United States*

E-mails: *jmbowma@emory.edu*; *dral@xmu.edu.cn*

[†]*Equal contribution*

**Abstract**

Machine learning potentials (MLPs) are widely applied as an efficient alternative way to represent potential energy surfaces (PES) in many chemical simulations. The MLPs are often evaluated with the root-mean-square errors on the test set drawn from the same distribution as the training data. Here, we systematically investigate the relationship between such test errors and the simulation accuracy with MLPs on an example of a full-dimensional, global PES for the glycine amino acid. Our results show that the errors in the test set do not unambiguously reflect the MLP performance in different simulation tasks such as relative conformer energies, barriers, vibrational levels, and zero-point vibrational energies. We also offer an easily accessible solution for improving the MLP quality in a simulation-oriented manner, yielding the most precise relative conformer energies and barriers. This solution also passed the stringent test by the diffusion Monte Carlo simulations.

**TOC Graphic**

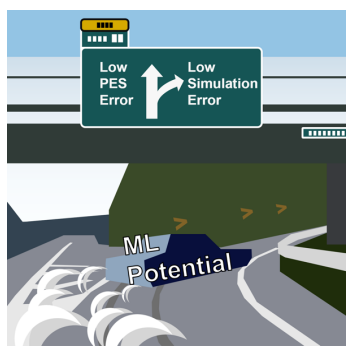

Potential energy surface (PES)[1-5] is one of the most essential concepts in computational chemistry, which can be abstracted as a multivariable function of the positions of nuclei with the output indicating the potential energy profile[6,7]. The negative gradients of this function are the forces that atoms experience. PES is a potent instrument that allows us to understand molecular structures[8], their stability[9-11], and reactivity governed by nuclear dynamics[12-18]. PES is extensively utilized for the exploration of the conformational landscape and reaction mechanisms[19-34].

Despite PES being an essential prerequisite to perform many chemical simulations, the construction often becomes a bottleneck in the application: highly accurate *ab initio* electronic structure methods offer reliable potential energies but have high computationally cost and scaling, while low-level empirical or semiempirical methods are fast but less robust. Fortunately, this conflict can be tackled by applying machine learning (ML) techniques to return an analytic functional form that mimics the PES, resulting in what is called machine learning potentials (MLPs).[35-51] By serving as an efficient replacement for quantum mechanically generated potential energies (and gradients), MLPs have developed rapidly over the past decade and various algorithms emerged. MLPs based on linear regression,[49,52] using permutationally invariant polynomials,[50,51] neural networks (NN),[44,47,53] and kernel methods (KM)[48,54-58] have been broadly applied for representations of correct ab initio PES for atomistic simulations in chemistry, physics, and materials science.[35-40, 43-52]

However, the convenience of MLPs, a black-box statistical approximator, comes with an inborn drawback of lacking the physical model foundation, making it difficult to get an intuitive sense of its reliability. The usual practice is to chase numerical metrics such as RMSEs (root-mean-square error) and MAEs (mean absolute error) on the test set drawn from the same distribution as the training data to evaluate MLPs' performance in PES. The hope is that these metrics would reflect the MLP performance in the actual simulation. Nevertheless, these numerical metrics, on the one hand, are *not representative*: most of the metrics are single numbers that only give an overall description of the error, this "average accuracy" cannot regard all the nuances in individual PES regions. On the other hand, they can be *biased* or even deceptive since the values are highly dependent on the composition of the test data, unable to reflect the poor prediction of the ML model for those data not drawn from the same distribution as the training set.

To overcome this limitation and obtain more reliable MLPs on chemical simulations, the obvious way is to directly validate MLPs by performing and examining the correctness of simulations with MLPs. An increasing number of studies attempted to evaluate MLPs beyond the statistical errors like MAEs and RMSEs on the test sets. Ong *et al.* presented an evaluation of MLP models based on local environment descriptors (GAP, MTP, NNP, SNAP, and qSNAP) and made a noteworthy observation of the absence of a consistent relationship between the accuracy in energies or forces and the accuracy in chemical simulations.[59] For instance, the SNAP model demonstrates superior performance in accurately reproducing polymorphic energy differences across various systems, even though it exhibits significantly higher RMSEs in energies and forces. Similarly, despite the GAP model exhibiting relatively low RMSE in predicting energies and forces, it demonstrates the least accuracy in polymorphic energy differences. Unfortunately, they did not delve into the underlying causation. Tkatchenko *et al.* assessed the performance of four MLPs including BPNN, SchNet, GAP/SOAP, and sGDML on the PES of glycine and azobenzene molecules and to check the reliability of MLPs, MD simulations were employed to reproduce the transition processes of azobenzene.[60] Their analysis revealed the qualitative problems with MLPs in trajectories. Csanyi *et al.* carried out error evaluations for simulation tasks beyond test RMSE, such as normal-mode prediction and dihedral torsional profile prediction.[61] More than that, MLPs whose test errors are not that bad may lead to catastrophic, non-physical events in the simulations: the most common one is molecular dissociation which should not happen.[61-64] Bowman and co-workers have introduced the term "DMC-certified" to indicate that the MLP can be used in a rigorous, diffusion Monte Carlo (DMC) calculation of the zero-point vibrational energy and wavefunction.[65] These studies provided valuable insight that the evaluation of MLPs only on the test RMSE is insufficient for judging their quality in actual simulations.

One common feature of the challenging cases is the proper description of the rich conformational space. The situation should be even more complicated when the global PES is needed,[39, 55, 66] i.e., for accurate vibrational spectra and zero-point vibrational energy calculations, e.g., with DMC, where the potential is needed that can be applied for very different geometries, including highly distorted ones. Training such potentials is a challenge in itself as the energy distribution is much broader and the quality of the reference quantum mechanical data can be problematic too. These challenges were the focus of a recent Perspective article that specifically examined the limitations of the MD17 dataset compared to much more extensive datasets for a number of molecules including malonaldehyde and glycine.[65] The latter molecule is the subject of this paper. The suggested solutions to fitting global PES try to improve the MLPs accuracy for the more important low-energy regions with, e.g., weighting schemes (where higher weights are assigned to lower energy during the fitting) or by even ignoring higher-energy points above some cutoff during the hyperparameter tuning stage and further refinement with the self-correction procedures. Such solutions are still quite cumbersome and usually require manual experimentations with the choice of suitable weighting functions[67] or cutoffs[55, 67] to obtain reasonable results.

Inspired by the above challenges and proposed solutions, we investigate how to construct robust, 'DMC-certified', MLP for the global PES featuring multiple conformers and transitions between them. We take a challenging example glycine which has eight different conformers (Figure 1).[67] Even non-global glycine PES was highlighted as a challenge to different classes of the state-of-the-art MLPs, including the neural network (NN) and kernel methods with local and global descriptors.[60] No clear solution was suggested. Glycine global PES was successfully fit with the permutationally invariant polynomial (PIP) MLP only after the use of a simple weighting function and using high-energy cutoff to ignore energy gradient information for highly distorted structures during the fitting.[67] Hence, the question also remains, whether such complicated cases can even be described with the other off-the-shelf MLPs and, more importantly, whether a general solution to make it possible and easier is feasible. In this study, we show for ANI-type NNs that while off-the-shelf implementation indeed cannot be used, it is possible to systematically improve their performance to the state-of-the-art level (i.e., achieving accuracy on par with PIP). ANI is an abbreviation for ANAKIN-ME, which in turn stands for Accurate NeurAl networK engINe for Molecular Energies.[68] It employs a modified version of Behler and Parrinello symmetry functions[35] to generate single-atom atomic environment vectors for the representation of molecules. Despite ANI's known problems of using the non-global descriptor[69] and being less accurate than some of the other MLP types based on the test RMSE evaluations.[46, 70] We develop a protocol to fit such an NN potential semi-automatically by focusing on getting the simulation result right instead of the lower test RMSE. The key idea behind the protocol is telling the NN what parts of PESs are more important than others by adjusting the weights for the errors in training points so that the resulting simulation error is the smallest. In our case, we use the flexible energy weighting function which gives higher weights for the more important low-energy regions of PES. This approach should be general enough to be applied in similar cases (e.g., different NN potentials and applications).

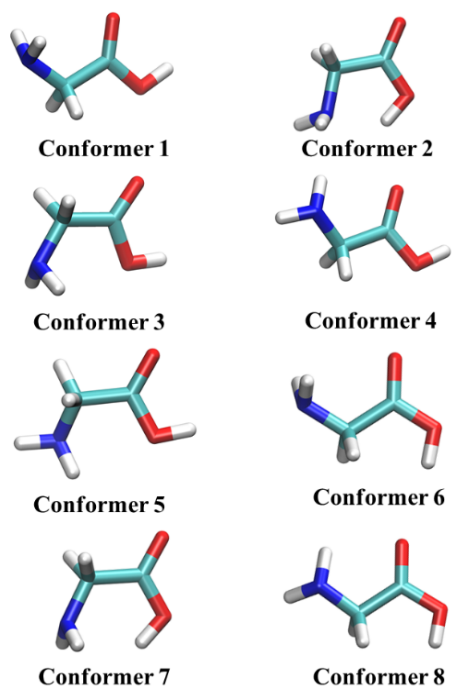

**Figure 1.** Equilibrium structures of eight conformers of glycine.

*How well does the test RMSE reflect the simulation performance?* We first investigate how well the test RMSE of an MLP reflects its performance for a glycine PES that aims to describe these conformers. The data set used is taken from the literature and its construction was not a focus of this study;[67] it contains 70099 geometries with B3LYP/aug-cc-pVDZ energies and gradients; these were shown to have excellent agreement with the gold-standard coupled-cluster level for the relative energies of the 8 conformers[71]. The dataset was obtained from every 10th step of AIMD trajectories initiated at the minima of conformers 1, 2, 3, 4 as well as several saddle points separating them for total energies between 100 and 80,000 cm$^{-1}$. Additional energies were added to fill in holes found in DMC calculations (details are given in Ref. [67]). To estimate the test RMSE we employ the standard approach of splitting the data set into the training and test set with a ratio of 9:1 and calculate the RMSE for energy predictions on the test set. We fit the ANI-type[53] NN on the training data set by using 10% of it for validation. The fitting is performed with the off-the-shelf setting as implemented in MLatom[72] interfaced to TorchANI[73] (see Methods). As the regions on the PES with very high energy are less important in actual simulations (see below), the general solution in the literature is to downweight the contributions from this region. Thus, here we also implemented the training of ANI-type NNs with a weighting function. We choose the weighting function which has the desired form of providing larger weights $w$ to points with lower energies while also having significant weights for higher-energy regions as defined by the energy $\Delta E$ (in hartree) relative to the global minimum:

$$w(\Delta E) = \max(\{[-6(a\Delta E)^5 + 15(a\Delta E)^4 - 10(a\Delta E)^3] + 1, \quad 0\}). \tag{1}$$

This function was the best among many possible choices we tried. The original form without parameter $a$ was introduced for a different problem in the literature.[74] The parameter $a$ gives us the additional flexibility of controlling how quickly the weighting function drops to 0 as the relative energy $\Delta E$ increases (Figure 2). The larger $a$ is, the more quickly the weighting function drops, and the less significant the high-energy data points are. Note that for $a = 0$, all weights $w = 1$, i.e., the training is equivalent to the off-the-shelf training of ANI without any weighting function. This is the only difference between the off-the-shelf ANI and other weighted variants we discuss below.

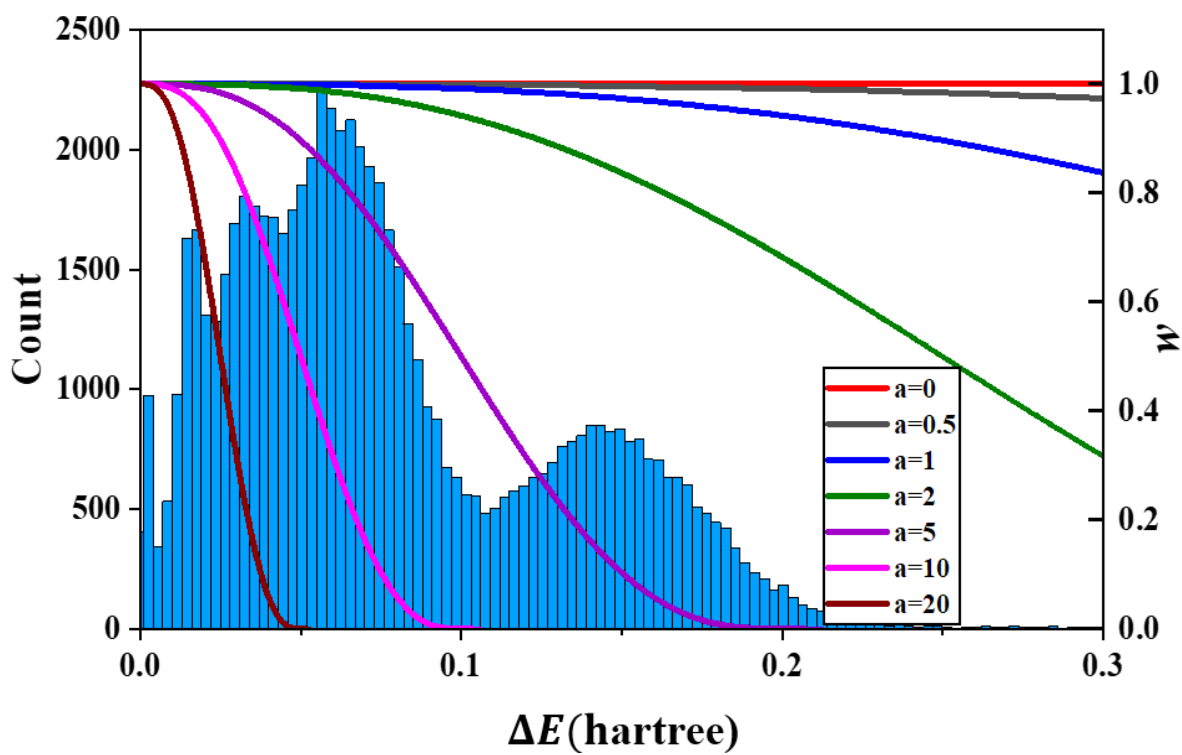

**Figure 2.** The shape of the weighting function is defined by Eq. 1 as the function of the energy $\Delta E$ relative to the global minimum. The dependence of the weighting function shape on the parameter $a$ is shown by the curves, and the histogram of the distribution of $\Delta E$ of the data set is also shown.

We train multiple MLPs for a set of parameters $a$ (including $a = 0$) by restarting the training with different randomly initiated NN weights, yielding different test RMSEs even for the same parameter $a$ (see Methods). For an illustration of the MLP's performance in simulations, we perform geometry optimizations of minima and transition states and evaluate the simulation error defined as the mean absolute error (MAE) in the energies of these stationary points relative to the global minimum. The general trend is that the lower the test RMSE the lower the simulation error (Figure 3). However, the clear correlation is only up to some point (denoted by the blue dotted line in Figure 3). For the region with smaller test RMSEs, no correlation is observed and the spread is quite broad. Two highlighted examples marked **A** and **B** in Figure 3 show extreme cases: for a relatively small test RMSE of 0.26 kcal/mol (point **A**) we can have a rather large simulation MAE of 0.14 kcal/mol for transition states while for a larger test RMSE of 0.36 kcal/mol (point **B**), we obtain two times smaller MAE of 0.08 kcal/mol. This has big implications if highly accurate results are needed as we cannot judge the quality of the MLP for the simulations solely based on the test RMSE. The simulation errors for the off-the-shelf ANI trained without the weighting function (equivalent to using $a = 0$ in the weighting function, marked 'unweighted' in Figure 3) are higher than the errors of both **A** and **B** models trained with the weighting function, while the test RMSE of off-the-shelf ANI is in between the test errors of the **A** and **B** models. All of these observations clearly illustrate the insufficient reliability of judging the model quality by only checking the test RMSE.

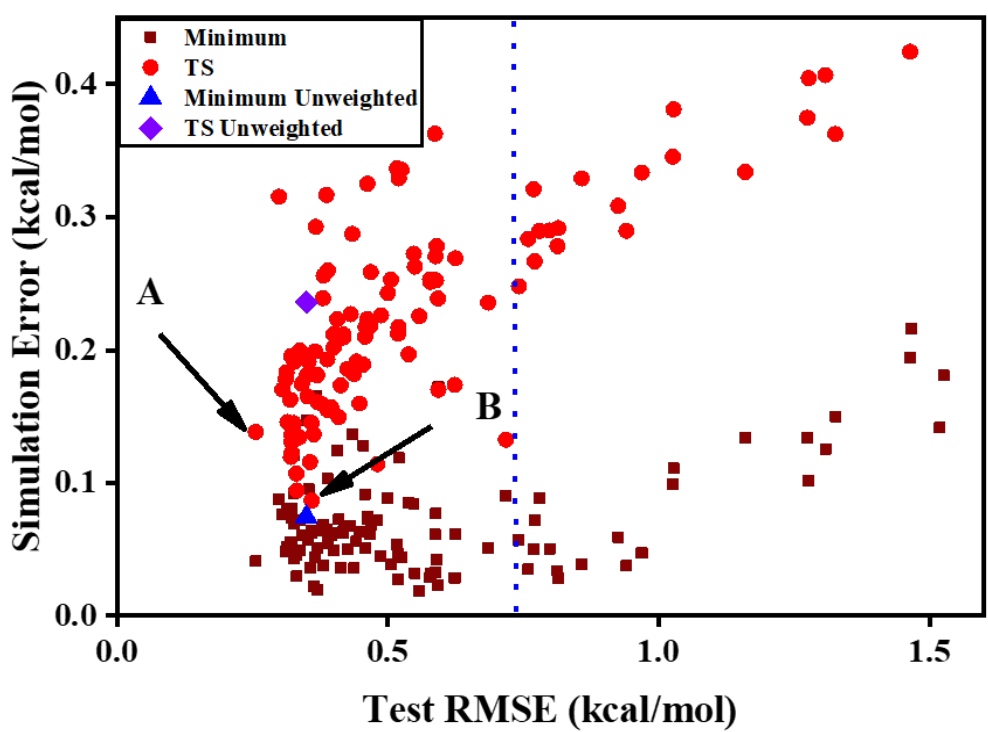

**Figure 3.** The relationship between the test RMSE and the simulation error (mean absolute error, MAE) performance of ANI-type NN in simulations (relative energies of conformers and transition states). Test RMSE is the root-mean-square error in energies, TS represents the MAE of MLPs for predicting fifteen main transition states connecting the local minima, Minima represents the MAE in predicting relative energies of all the eight minima compared with the B3LYP/aug-cc-pVDZ method.

*Simulation-oriented training strategy.* Above we saw that the simulation result is very sensitive to the MLP training and just looking at the test RMSE does not provide a confident estimate of the quality of the simulations. Thus, here we report the implementation of the automatable, simulation-oriented training protocol aiming at obtaining the robust MLP with good performance in actual simulations rather than only checking the test RMSE.

The key to successful simulation-oriented training lies in a) choosing an appropriate, computationally affordable, property-based metric describing the quality of the simulations and b) designing an effective model-optimization strategy based on the chosen metric. We fine-tune the weighting function to optimize the model so that it produces better performance in simulations judged property-based metric. Here, we choose the error in the relative energies of eight glycine conformers. To independently validate the model we use the related but different property-based metrics such as errors in transition state energies, normal mode frequencies, and zero-point vibrational energies.

Of course, the reference values for the property-based metric used for model optimization should be known. It means that the benefit of using MLP should be then in applying it to related but much more computationally expensive simulation tasks. Later we will show that the resulting MLP models can be successfully used for such related tasks: very computationally expensive DMC calculations.

To design an effective model-optimization strategy, we optimize the parameter *a* in the weighting function defined by Eq. 1 that controls the regional importance of the training points based on relative energies. We treat this parameter as the hyperparameter to be tuned for the model selection, i.e., we optimize it to obtain the lowest simulation MAE in relative energies of conformers with the MLP being fitted.

Model selection can be easily done given the simplicity of our approach: we just need to scan different values of the parameter *a* to obtain the lowest simulation MAE for the minima. We see that such an optimal value of *a* is around 2 for the global glycine PES (Figure 4).

To ensure that this strategy does not overfit, we also investigate whether lower errors for one type of simulation lead to lower errors in another, related, type of simulation. Here, we check the MAE for the relative energy values of fifteen primary transition states predicted by the ANI (denoted as TS, red circles in Figure 4). These MAEs were not used for the model selection, i.e., they represent a completely independent test. The trend observed in the TS convincingly aligns with the trend observed for the minima. This indicates a close correlation between the accuracy of transition states predicted by the MLP and the accuracy of the energy minima predicted by the same MLP trained using our strategy. Furthermore, it is noteworthy that as the parameter increases (exceeding 2.5), the TS energies show a more rapid increase in errors, which implies that while different simulations (such as identifying energy minima and transition states on the PES) exhibit an overall similar dependence on the weighting function, they do not share the same level of sensitivity to different regions in the training PES data set.

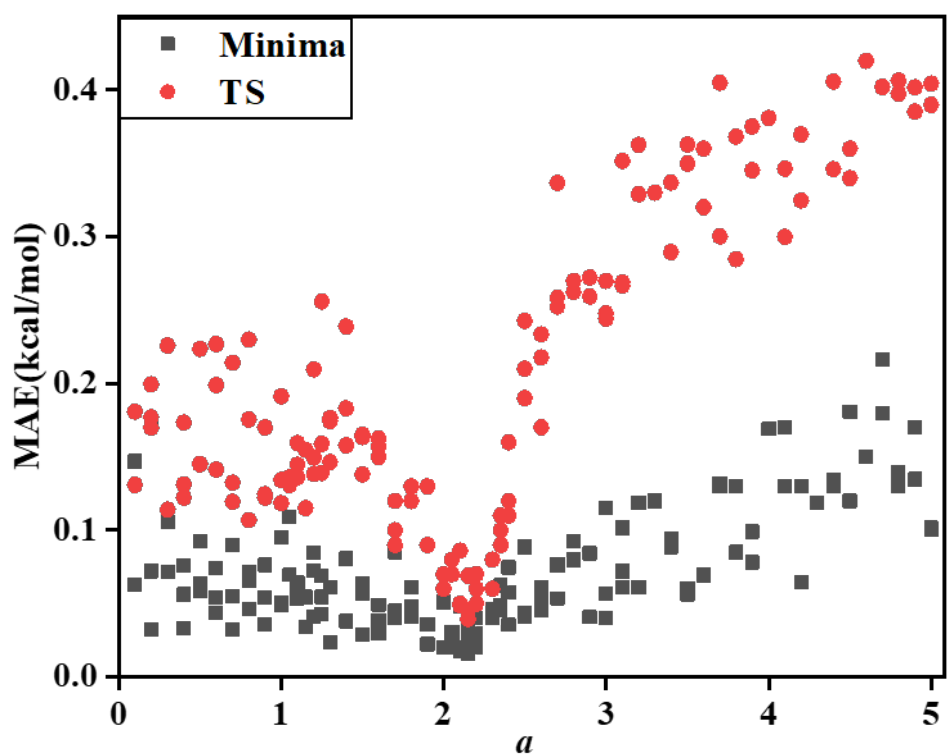

**Figure 4.** Optimization of the parameter *a* in the weighting function. MAE represents the mean absolute error of MLPs for predicting relative energies of all the eight minima compared with B3LYP/aug-cc-pVDZ method (labeled Minima, black) and of the fifteen main transition states (labeled TS, red) connecting the local minima.

*Performance for the minima and transition states*. Having identified the optimal range of *a* around 2.0 that corresponds to both small MAE in relative energies of minima and TSs, we subsequently fine-tuned *a* around 2.0 based on MAE in minima and ultimately obtained the best model with *a* set to 2.15. We will refer to this MLP as 'the best ANI model'. The best ANI model has an MAE value of 0.016 kcal/mol for the relative energies of the eight minima and an MAE value of 0.039 kcal/mol for the relative energies of the 15 primary saddle points (Table 1). The simulation errors of energy-weighted ANI models with *a* set to 2.15 are consistently much lower than those of off-the-shelf ANI trained without using the weighting function (as can be seen from results of repeated experiments, Table 1). This is despite off-the-shelf ANI model's lower test RMSEs indicating that the lower test RMSEs do not guarantee better performance in simulations.

The energies of each conformer relative to the global minimum given by the best ANI model and the reference B3LYP/aug-cc-pVDZ method are shown in Figure 5. Since all the points are essentially located on the diagonal of Figure 5, this implies the best ANI model we trained under the energy-weighted scheme is capable of accurately identifying all the eight energy minima and fifteen main saddle points on the PES of glycine. Compared to the B3LYP/aug-cc-pVDZ method which generates the dataset, our training can significantly reduce the computational cost for simulations without sacrificing accuracy.

**Table 1.** Statistics of simulation errors and test RMSE in kcal/mol with the off-the-shelf ANI and the ANI trained with the optimized weighting parameter $a$= 2.15 (energy-weighted ANI). Each set of models is trained five times. Best ANI and PIP are single models, where the best ANI is the best performing energy-weighted ANI (with the lowest MAE in minima).

|  | off-the-shelf ANI | energy-weighted ANI | best ANI | PIP[67] |
|---|---|---|---|---|
| MAE in minima | 0.079±0.024 | 0.023±0.0062 | 0.016 | 0.021 |
| MAE in TS | 0.19±0.040 | 0.055±0.0088 | 0.039 | 0.35 |
| test RMSE | 0.38±0.077 | 0.41±0.036 | 0.45 | – |

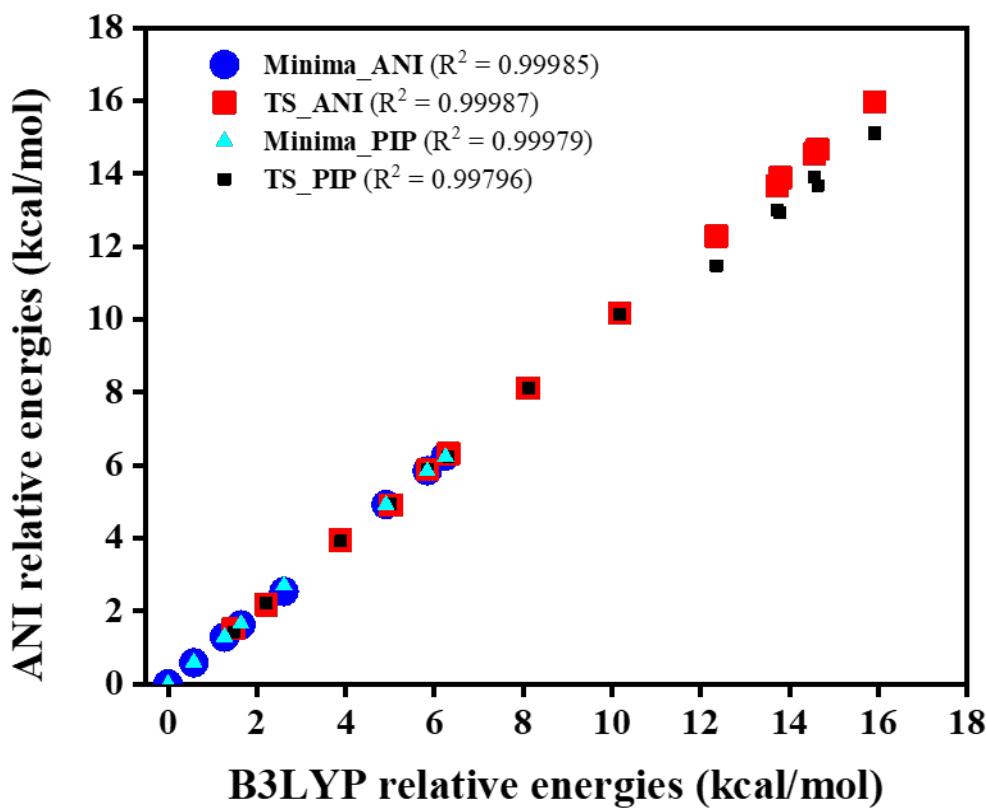

**Figure 5.** Energies relative to the global minimum of conformers and saddle points obtained by the best ANI model, PIP potential and the B3LYP/aug-cc-pVDZ.

*Performance for the vibrational levels.* To further ensure the reliability of the best ANI model we obtained, we subsequently performed normal-mode calculations for the eight conformers and the same fifteen saddle points that were previously investigated. For each optimized geometry, the twenty-four harmonic frequencies were calculated using the reference DFT method (B3LYP/aug-cc-pVDZ) and the best ANI model, and MAE was calculated based on the deviation between ANI and DFT results. The MAEs of harmonic frequencies of the minima and transition states are shown in Figure 6. The best ANI model is evidently the most robust MLP with the MAEs of only a few wavenumbers for the energy minima. Impressively, ANI potential can achieve good agreement not only for transition states in the low-energy region but also for those in the high-energy region (16, 23, 25, 37, 57, and 48 in Figure 6), whose energies are higher than 11 kcal/mol. The PIP has an overall similar performance for low-energy TSs but is markedly worse for the higher-energy TS which might be because its training used a different training approach (another weighting function, removed energy gradients for higher-energy regions, and no optimization of the weighting function).

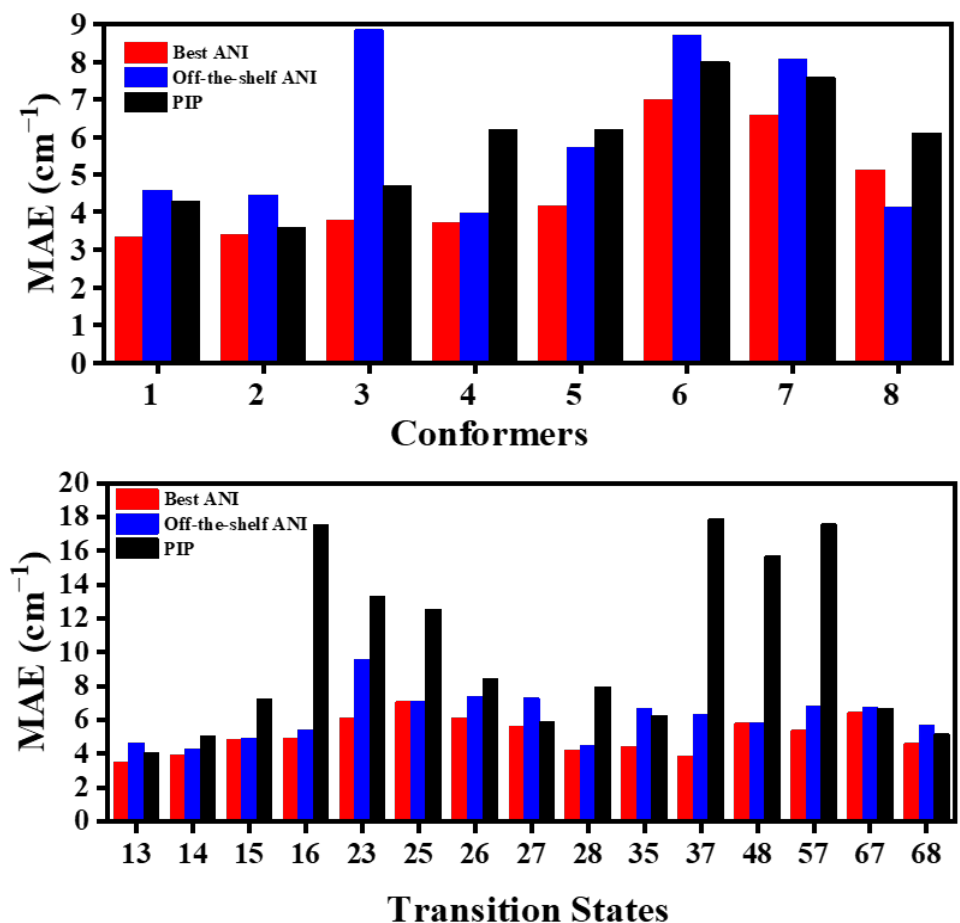

**Figure 6.** Mean absolute errors (MAE, in cm$^{-1}$) for the harmonic frequencies of energy minima and transition states (their selection and numbering taken from literature[67]) obtained from the ANI potential (red), the off-the-shelf ANI potential (blue) and PIP potential (black) compared to those at B3LYP/aug-cc-pVDZ level.

*Performance for the harmonic zero-point vibrational energies (ZPVE).* Based on the harmonic frequencies we calculated for energy minima and transition states, we further calculated their approximate harmonic zero-point vibrational energies (ZPVEs); the results are shown in Table **2**. For the off-the-shelf ANI model, having a smaller test RMSE cannot guarantee its accuracy in predicting better harmonic ZPVEs. For the transition states, this is especially evident when compared to the performance of the best ANI model, which achieved nearly half of the error. This result highlights that solely relying on the test error may be deceptive.

**Table 2**. Statistics of simulation errors (MAE in harmonic zero-point vibrational energies for minima and transition states in cm$^{-1}$) and test RMSE in kcal/mol with one of the off-the-shelf ANI models, the best ANI trained with the energy-weighing function and optimized parameter $a$= 2.15, and PIP.

|  | off-the-shelf ANI | best ANI | PIP[67] |
|---|---|---|---|
| MAE in minima | 33.9 cm$^{-1}$ | 24.8 cm$^{-1}$ | 27.6 cm$^{-1}$ |
| MAE in TS | 42.0 cm$^{-1}$ | 15.0 cm$^{-1}$ | 30.4 cm$^{-1}$ |
| test RMSE | 0.29 kcal/mol | 0.45 kcal/mol | – |

*Performance for the DMC-level zero-point vibrational energies (ZPVE)*. To give a more comprehensive description of the morphology around the local minima of the MLP, we calculate the zero-point vibrational energies (ZPVE) of each conformer employing rigorous diffusion Monte Carlo (DMC) simulations. The results are presented in Table **3**. The agreement between the best ANI model and previously reported state-of-the-art PIP potential is remarkable taking into account that DMC samples have a much larger space than harmonic-oscillator frequencies (we do not have reference B3LYP calculations due to their prohibitive cost for such simulations). Most importantly, ANI can correctly describe the key feature of the double-well pairs on PES (conformers 1 and 4, 2 and 7, 3 and 5, and 6 and 8) as the ZPVEs for the conformers in each pair are very close. The differences between ZPVEs for each conformer in the double-well pair range from 1 $cm^{-1}$ to 14 $cm^{-1}$ (0.003 to 0.04 kcal/mol). They are somewhat larger than the differences with PIP (1–3 $cm^{-1}$), however, the latter were averaged over several repeated DMC calculations while ANI results are bootstrapped from a single repeat which overall leads to higher standard deviations in ANI's ZPVEs and may partly explain the overall better performance of PIP. The results discussed above ensure the reliability of the MLP constructed by the best ANI as they are 'DMC-certified'. It also demonstrates the high quality of the PES data set.

Unfortunately, we cannot compare ANI or PIP results with thye reference calculations as they would require the DMC simulations with the B3LYP/aug-cc-pVDZ method for all eight conformers. One DMC simulation for a single conformer requires 30000 walkers and 55000 steps comprising roughly $1.6 \cdot 10^9$ potential evaluations with B3LYP. This is the reason why the MLPs (PIP or ANI) are fitted in the first place – to enable expensive calculations such as DMC or *ab initio* MD.

**Table 3.** Zero-point vibrational energies (in cm$^{-1}$) of each conformer, obtained from DMC by ANI and PIP. Note that PIP results are averages of several simulation repeats while ANI results are bootstrapped from a single repeat leading to higher standard deviations.

| Conformer | Best ANI | PIP[67] |
| --- | --- | --- |
| 1 | 17154 ± 38 | 17151 ± 5 |
| 4 | 17140 ± 20 | 17150 ± 6 |
| 2 | 17442 ± 40 | 17430 ± 3 |
| 7 | 17434 ± 34 | 17430 ± 4 |
| 3 | 17701 ± 27 | 17717 ± 4 |
| 5 | 17696 ± 34 | 17720 ± 6 |
| 6 | 18819 ± 24 | 18843 ± 6 |
| 8 | 18818 ± 31 | 18841 ± 4 |

*How does the simulation-oriented training of ANI compare to the state-of-the-art PIP?* The focus of this paper was to investigate how to properly train popular MLPs on challenging data sets. We demonstrated that special care is needed for a balanced description of different PES regions which we achieve by introducing the protocol based on the energy-weighted function whose shape is adjusted to produce the lowest error in the computationally cheap simulation. This MLP can then be used for much more computationally expensive calculations such as DMC.

Although comparison with other MLPs is not the focus, it is important to discuss the results in the context of the previous state-of-the-art PIP calculations. In this study, we knew the PIP results and hence had the opportunity to strive to achieve better performance than PIP. However, it is important to emphasize the published PIP results used different weighting functions which were not optimized using the protocol introduced here. Hence, our results with ANI can be viewed as additional, independent validation of the PIP fit on glycine: our best ANI results agree very well with PIP for most of the examined simulation errors (Table 1–

**Table 3** and Figure 5–Figure **6**). The only exception is high-energy TSs, which are described better by ANI. These TSs are less important for simulations with DMC and the higher MAE with PIP can be explained by using another weighting function which goes much steeper and faster to zero in the high-energy region and also the neglect of the gradients for high-energy geometries. In the future, PIP will likely adopt a similar simulation-oriented protocol for finding the best weighting function shape.

Finally, given the near-perfect agreement between the DMC zero-point vibrational energies as well as the excellent agreement for the relative energies and harmonic frequencies of the minima it is worth commenting on the speed of evaluation. This is especially relevant for DMC calculations, which are highly compute intensive. Previous comparisons of ANI and PIP for similarly precise fits for ethanol indicate that the PIP PES runs roughly 10–100 times faster than the ANI PES based on the evaluation with the old version of MLatom.[75] This factor is consistent with a more recent assessment of speed and precision for ANI and PIP PESs for aspirin.[76]

For glycine, running 30000 walkers for DMC simulations (with 55000 steps in each walker) with ANI took ca. 60 hours on 1 NVIDIA 3090 GPU, and would take ca. 107 CPU-hours on Intel(R) Xeon(R) Gold 6240 CPU @2.60GHz processors with the new MLatom version with improved performance. The published timings for PIP evaluation for the same number of walkers and steps on a similar Xeon processor (2.4 GHz Intel) took roughly 30 CPU-hours;[67] a factor of ca. 3.5 faster (based on CPU-time comparison). The training speed is also a factor. Training of ANI on 1 NVIDIA 3090 GPU took ca. 48 h, while PIP took only 12.5 hours (3.5 hours to obtain the PIP basis and 9 hours for the fit) on a single CPUs of the above hardware types.

Performance improvements of both ANI and PIP methods are certainly possible. ANI evaluations can be sped up by, e.g., converting the network in the C++ code and recompiling it into a more efficient executable and by reducing the size of the network which is probably unnecessarily large for 70k training points. (The ANI network was designed for training on millions of points but we did not optimize its architecture for this specific application to keep the modifications to the default settings of ANI to the minimum.) Performance enhancements of PIPs are also possible. One that has already been reported for the PIP PES for aspirin[76] is the reduction of the PIP basis with very little loss in precision but a significant speed-up in both training and prediction.

*Conclusions and outlook.* We demonstrate the performance of the off-the-shelf machine learning potentials cannot be solely judged by the test RMSEs but rather should be evaluated on the simulation errors.

To systematically improve the quality of MLPs, we propose simulation-oriented training which utilizes the energy-weighted training scheme where a single hyperparameter is tuned to obtain the better performance of MLPs in simulations. We demonstrate the underlying relationship between various simulations, wherein the precision in predicting relative energies for energy minima influences the accuracy in predicting transition states. Importantly, we can identify a region in the parameter space where both the simulation accuracy for the energy minima and transition states are high and obtain the best MLP model around that region.

The best MLP model's reliability is further confirmed through normal-mode calculations for energy minima and transition states. Additionally, the advanced DMC method is utilized to obtain zero-point vibrational energies (ZPVEs) for energy minima. All the simulation results obtained ensure the reliability of the MLP obtained through a simulation-oriented approach.

To facilitate the usage of the method, we provide a convenient tool implemented in MLatom, an open-source package that can run both locally and on the online XACS cloud computing.[77]

While the focus of this study was how we can properly train the model given already sampled points, a promising direction for future investigation is combining different sampling procedures such as active learning[78-82] (and its often-used variant adaptive sampling[83]) with our simulation-oriented training strategy.

**Methods**

*ANI model.* A development version of MLatom[72] was used in this work. Its interface to TorchANI[73] (version 2.2.3) was used for generating the AEV descriptor used in ANI models. The parameters of the AEV descriptor were chosen from the ANI-1ccx[84] model. Each ANI model consists of four networks for the H/C/N/O atoms, respectively. While each network implements 3 hidden layers of 160, 128, and 96 CELU-activated neurons.

In the training process, both losses on energy ($\mathcal{L}_E$) and gradients ($\mathcal{L}_{grad}$) are calculated with mean squared error weighted by relative energy reference and accounted for in the final loss $\mathcal{L}$:

$$\mathcal{L} = \mathcal{L}_\text{E} + 0.1 \cdot \mathcal{L}_\text{grad} = \frac{\sum_i^{N_\text{tr}} \left(w_i(\hat{E}_i - E_i)\right)^2}{N_\text{tr}} + 0.1 \cdot \frac{\sum_i^{N_\text{tr}} \sum_j^{3N_\text{at}} \left(w_i(\hat{F}_{i,j} - F_{i,j})\right)^2}{3N_\text{at} N_\text{tr}}, \quad (2)$$

where $w_i$ is the weight calculated with Eq. 1, $E_i$ is the centered energy of the training point $i$, and the $F_{i,j}$ is the force component of the training point $i$ (negative energy gradient).

The learning rate was scheduled by the *ReduceLROnPlateau* scheduler provided by PyTorch[85], monitoring $\mathcal{L}_\text{E}$ with patience of 32 epochs and a reduce factor of 0.5 to reduce it from the initial value of $1 \times 10^{-3}$. Early stopping triggered by a minimum learning rate of $1 \times 10^{-5}$ was also used to prevent overfitting.

The tuning process of parameter *a* starts from 0.1 and was performed at intervals of 0.1, continuing until it reached 5, with 3 repeats at each point. While the unweighted, off-the-shelf ANI was trained for 5 times.

*Geometry optimization*. Geometry optimization on ANI potentials was performed with the same version of MLatom as above. The optimization algorithm was performed via the interface to Gaussian 16[86]. Optimization settings used in Gaussian were *opt(nomicro,calcall,noeigen)* for energy minima, and *opt(ts,calcfc,noeigen,nomicro)* for transition states. The optimized geometries at B3LYP/ aug-cc-pVDZ reported in the same literature[67] of the glycine data set were used as the initial guesses.

*Diffusion Monte Carlo*. Diffusion Monte Carlo calculations for ANI potentials were performed by *PyVibDMC* module (https://github.com/rjdirisio/pyvibdmc, version 1.3.7) interfaced to MLatom. For faster DMC calculations, we implemented in MLatom an efficient batch parallelization of the potential surface evaluations from many walkers. For each energy minimum, the pre-optimized geometry was scaled by a factor of 1.01 then was used as the starting geometry for 30000 random walkers. Every simulation consists of 55000 steps of 1 atomic unit of time each, and the zero-point vibrational energy was bootstrapped by averaging the mean energies from eight segments with 5000 steps after the 15000[th] step, and a standard deviation of the mean values was calculated from these eight segments. The convergence of each simulation was manually checked.

**Data availability**

No data was generated in this study.

## Code availability

The code for all simulations is publicly available in the open-source package MLatom (https://github.com/dralgroup/mlatom). Best, off-the-shelf, and **A** and **B** examples of trained ANI potentials are available at https://github.com/dralgroup/ani_glycine24. This repository also describes how to train the models with and without the weighting function and use them with MLatom for the simulations. The PIP PES is available online as supplementary material at https://pubs.aip.org/aip/jcp/article/153/24/244301/200351/Full-dimensional-ab-initio-potential-energy.

## Author contributions

J.M.B. and P.O.D. initiated the study and outlined the initial study plan. F.G. made implementations and R.W. performed a selection of energy-weighting functions, their parameters, and calculations, both analyzed the calculations and wrote the manuscript under the supervision of P.O.D. P.Z. contributed to the original implementation of the geometry optimizations. A.N. and C.Q. provided additional explanations about the PIP calculations. All authors discussed the project and contributed to the final version of the manuscript.

## Acknowledgments

P.O.D. acknowledges funding by the National Natural Science Foundation of China (No. 22003051 and funding via the Outstanding Youth Scholars (Overseas, 2021) project), the Fundamental Research Funds for the Central Universities (No. 20720210092). This project is supported by Science and Technology Projects of Innovation Laboratory for Sciences and Technologies of Energy Materials of Fujian Province (IKKEM) (No: RD2022070103). R.C. thanks Università degli Studi di Milano for funding under grant PSR2022_DIP_005_PI_RCONT. J.B. acknowledges support from NASA grant (80NSSC22K1167) The authors also thank Adrian Roitberg for the initial discussions on off-the-shelf ANI. R.W. thanks Shengheng Yan for the discussions and his help with technical issues.